\documentclass[manuscript]{acmart} 

\AtBeginDocument{%
  \providecommand\BibTeX{{%
    \normalfont B\kern-0.5em{\scshape i\kern-0.25em b}\kern-0.8em\TeX}}}

\setcopyright{acmcopyright}
\copyrightyear{2023}
\acmYear{2023}
\acmDOI{XXXXXXX.XXXXXXX}

\begin{document}

\title{PokAR: Facilitating Poker Play Through Augmented Reality}

\author{Adam Gamba}
\email{agamba@princeton.edu}
\author{Andrés Monroy-Hernández}
\email{andresmh@princeton.edu }
\affiliation{%
  \institution{Princeton University}
  \city{Princeton}
  \state{New Jersey}
  \country{USA}
  \postcode{08544}
}

\renewcommand{\shortauthors}{Gamba and Monroy-Hernández}
\begin{abstract}
We introduce PokAR, an augmented reality (AR) application to facilitate poker play. PokAR aims to alleviate three difficulties of traditional poker by leveraging AR technology: (1) need to have physical poker chips, (2) complex rules of poker, (3) slow game pace caused by laborious tasks. Despite the potential benefits of AR in poker, not much research has been done in the field. In fact, PokAR is the first application to enable AR poker on a mobile device without requiring extra costly equipment. This has been done by creating a Snapchat Lens \footnote{A Lens in Snapchat is an experience that utilizes augmented reality to transform the world around you \cite{snap_how_nodate}.} which can be used on most mobile devices. We evaluated this application by instructing 4 participant dyads to use PokAR to engage in poker play and respond to survey questions about their experience. We found that most PokAR features were positively received, AR did not significantly improve nor hinder socialization, PokAR slightly increased the game pace, and participants had an overall enjoyable experience with the Lens. These findings led to three major conclusions: (1) AR has the potential to augment and simplify traditional table games, (2) AR should not be used to replace traditional experiences, only augment them, (3) Future work includes additional features like increased tactility and statistical annotations.
\end{abstract}


\ccsdesc[300]{Human-centered computing~Collaborative and social computing devices}

\keywords{connected lens, augmented reality, poker, co-located, interaction, socialization}

\begin{teaserfigure}
\centering
  \includegraphics[width=350pt]{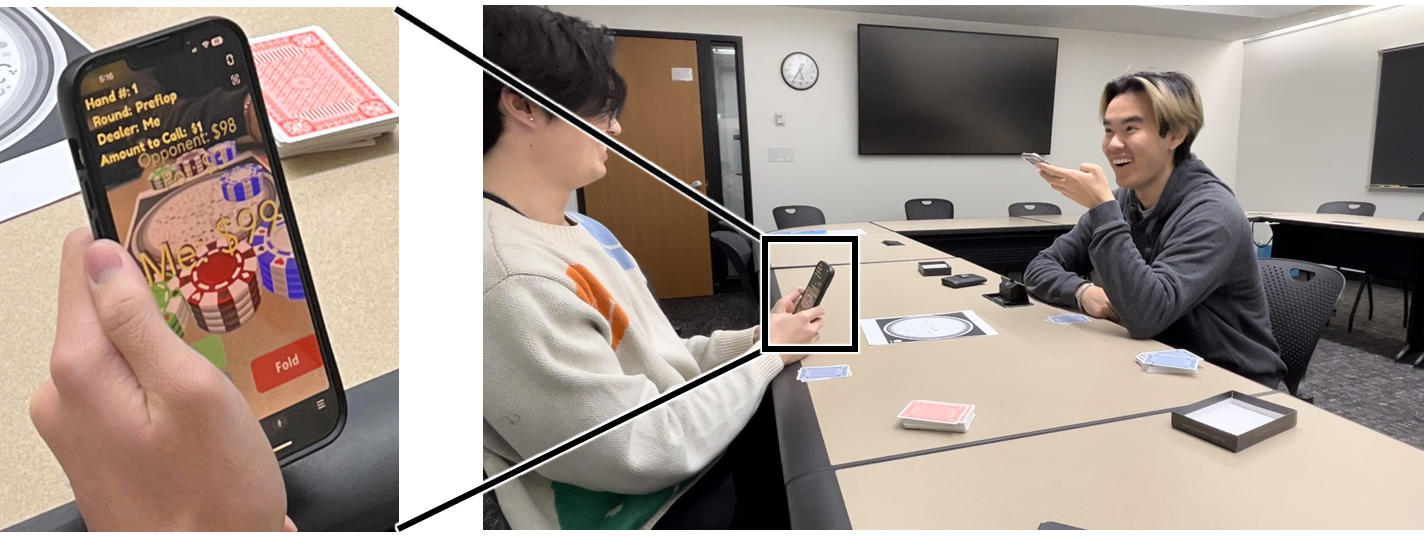}
  \caption{Two players using the PokAR application.}
  \label{fig:teaser}
\end{teaserfigure}



\maketitle

\section{Introduction}
\hspace{\parindent}
The goal of this project is to facilitate heads-up Texas hold’em poker play through augmented reality. Poker is cumbersome to play in its current form, requiring players to have poker chips and knowledge of the complex rules to play correctly. Without an experienced player to guide the game, new players often find it difficult to learn the rules and play correctly \cite{sakuma_enhancing_2012}. Additionally, due to the burden of physical chips, it is difficult to play poker in many scenarios (e.g., at the beach, while traveling, or camping). Finally, the game pace is often slowed due to poker's complex rules and the need for laborious tasks like counting chip stacks.

Augmented reality technology is well-equipped to solve these issues in three ways. Firstly, AR can eliminate the need for physical poker chips by instead utilizing AR to render chips. Next, AR can help guide players through the complex rules of poker by hinting at legal actions during gameplay. Finally, AR can help decrease the burden of laborious tasks (like counting chips) and increase the game pace. PokAR helps alleviate these three issues, which we’ll discuss further throughout this paper.

Poker is a popular game, with over 120 million players worldwide playing regularly online \cite{fast_offshore_online_2021}. Texas hold'em is one of the most popular poker variants. In this variant, players are dealt two private cards and five community cards, and they battle to make the best hand or bluff opponents into folding. 'Heads-up' poker is a term used to describe poker played by just two players, head-to-head. In its current state, PokAR supports only heads-up Texas hold'em poker, but with future work, it could be extended to more players and more variants. Throughout this paper, we will use the term 'poker' to refer to heads-up Texas hold'em poker. 

Poker is a classic example of a social, co-located game, since poker, by design, emphasizes in-person, co-located interaction. Players often look at each other and speak to each other during a poker game, either to gain information or to socialize. Also, poker forces players to focus on the same enablers, or "physical objects that trigger and are the focus of the AR experience" \cite{dagan_project_2022}. These enablers, like playing cards and poker chips, can help guide an AR experience and engage players more closely than in games that do not have a similar shared focus. For the above reasons, we chose to augment poker in this study.

PokAR is not intended to replace traditional poker, rather it helps people play when traditional poker would be difficult or impractical. The goal of AR applications should be to disappear completely and seamlessly immerse the user in a realistic experience that combines reality with augmentation \cite{weiser_computer_1991}. This disappearance frees users to utilize these applications more effortlessly, allowing them to focus on new goals, beyond the application itself.

\section{Related Work}
\hspace{\parindent}
While we have established AR as a possible solution to the aforementioned issues with poker, very little work has been done concerning AR poker. Additionally, while AR is not yet a heavily explored area, researchers have argued that “A Poker-Assistance-Software is an ideal test area for an AR Application with real added value,” with possible areas to add value including automation and statistical estimations \cite{thul_pokertool_2013}.

Similar projects in the past have all relied on physical means to augment reality. For example, researchers used overhead projectors to project all aspects of the poker game (e.g., cards, chips, etc.) onto a table \cite{sakuma_enhancing_2012}. Additionally, researchers have used RFID playing cards to detect the dealt cards \cite{sakuma_enhancing_2012}. While this study succeeded in creating an AR application to ease some of the same cumbersome aspects of poker tackled by PokAR (physical chips, complex rules, slow game pace), it did so using a high-cost solution, which is impractical for most recreational use. Additionally, they disregarded studying how this AR setup influenced social interactions in poker.

Furthermore, people have utilized virtual reality (VR) in the past to create commercial poker video games. One example is PokerVR by Meta, which uses “expressive avatars built for reading tells with growing customizations” \cite{meta_poker_2019}. While they may say this, VR poker applications still just employ static players' avatars, which do not emphasize the in-person, social nature of poker.

PokAR is the first project to enable AR poker without needing additional equipment other than a mobile device and a regular deck of cards (like an overhead projector or a VR headset). This is a worthwhile problem because it is the first application to enable AR poker at a low cost, since it only requires a few commonly-owned pieces of equipment (mobile devices and playing cards). Additionally, utilizing AR over VR allows the gameplay to emphasize the social aspects of co-location and increase socialization when compared to VR implementations.

Co-located gaming has been shown to lead to more effective and enjoyable gaming, as players can more easily communicate and build social relationships \cite{herodotou_iterative_2015}. One major reason for this is "out-of-the-game, game-related communication" \cite{herodotou_social_2010}. By having the ability to converse about other topics while simultaneously being involved in a game with another player, these players are given the opportunity to build a deeper connection.

Additionally, when comparing socialization in AR and VR, prior research tends to support increased socialization in AR applications. AR games have the ability to "potentially enhance social communication and social interaction between people" \cite{savela_does_2020}, whereas high-involvement in VR games could potentially isolate users socially and "negatively affect their well-being" \cite{lee_social_2021}. Thus, we chose to develop an AR application rather than a VR application to reap the social benefits of shared, co-located experiences. 

\section{PokAR System}
\hspace{\parindent}
The PokAR Snapchat Lens allows users to play heads-up poker with another player on two mobile devices. AR visual annotations include 3D models of poker chips which dynamically render with changing stack size, and 3D text above the chip stacks denoting the size of each stack. 2D visual annotations include number of hands played, current round, current dealer, previous action, amount to call, waiting message, and UI buttons with labels "Check," "Call," "Bet," "Raise," and "Fold." All AR and 2D annotations render dynamically with changing game state, stack amounts, and legal actions. PokAR implements five main features to help achieve its motivating goals.

\begin{itemize}
    \item “3D AR Chips” - PokAR renders 3D models of chips to eliminate the requirement of needing physical poker chips.
    \item “UI Action Buttons” - 2D buttons rendered on the player’s screen allows them to select among and perform legal actions at the current state of the game.
    \item “Game Messages” - Messages provide additional information to both players about the actions of players, bet amounts, and more throughout the game.
    \item “Counting Stacks” - A live count of the number of chips in all chip stacks is rendered above the 3D models, eliminating the hassle of counting chips manually.
    \item “Awarding Pots” - After a winner is determined (through folding or at showdown), the chips are automatically awarded to the winner, eliminating the hassle of moving chips manually.
\end{itemize}
\subsection{Approach}
\begin{figure}
  \centering
  \includegraphics[width=200pt]{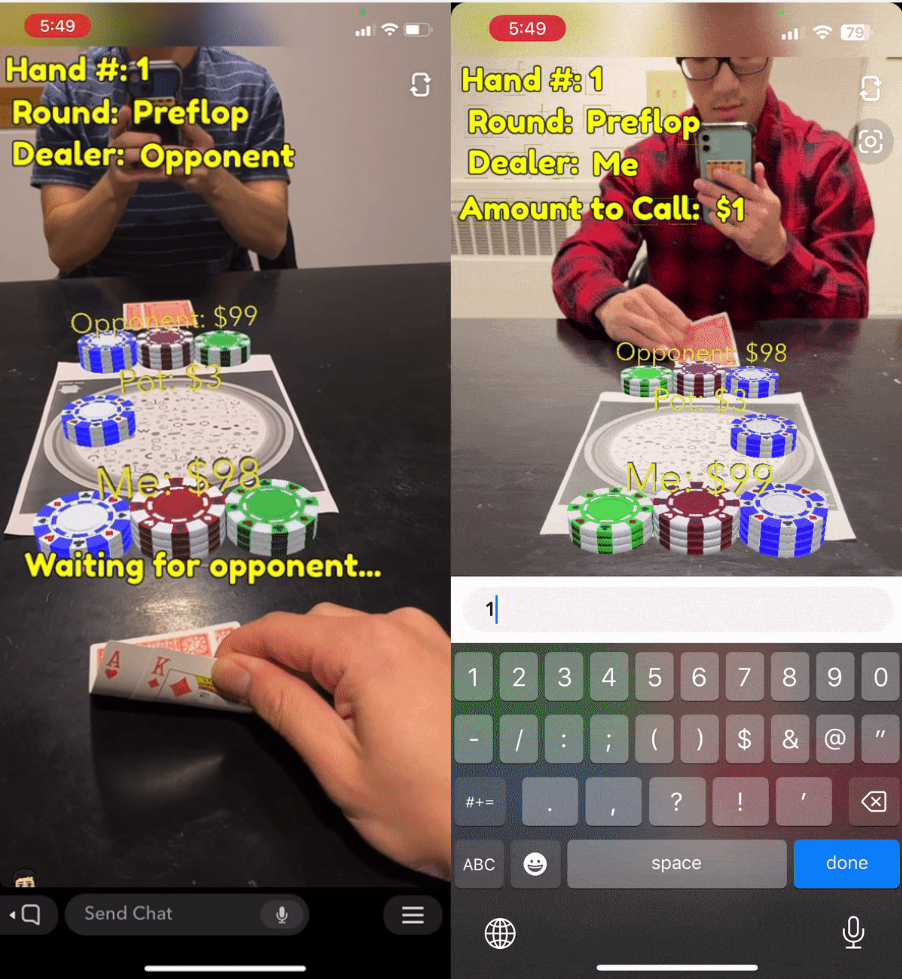}
  \caption{Side-by-side points of view of the same game of PokAR on two mobile devices.}
  \label{fig:1st-person}
\end{figure}

\hspace{\parindent}
To achieve our motivating goals, we developed a Snapchat Lens which could be used on any mobile device to facilitate poker play through AR. Besides having physical playing cards, players will play the game of poker in AR by interacting with their augmented environment through a mobile device. We used the Snap Lens Studio IDE with JavaScript for development, the Snapchat app for testing and deployment, and GitHub for version control. Code for this project can be found at the link in Appendix D. Physical playing cards act as an enabler to ground the game in physical reality. A demonstration of the completed application is shown in Fig~\ref{fig:1st-person}. In the development of PokAR, we faced and solved three major implementation subproblems. They will be discussed below.

\subsection{Subproblem 1: Modeling Poker in Code}
\begin{figure}
  \centering
  \includegraphics[width=\linewidth]{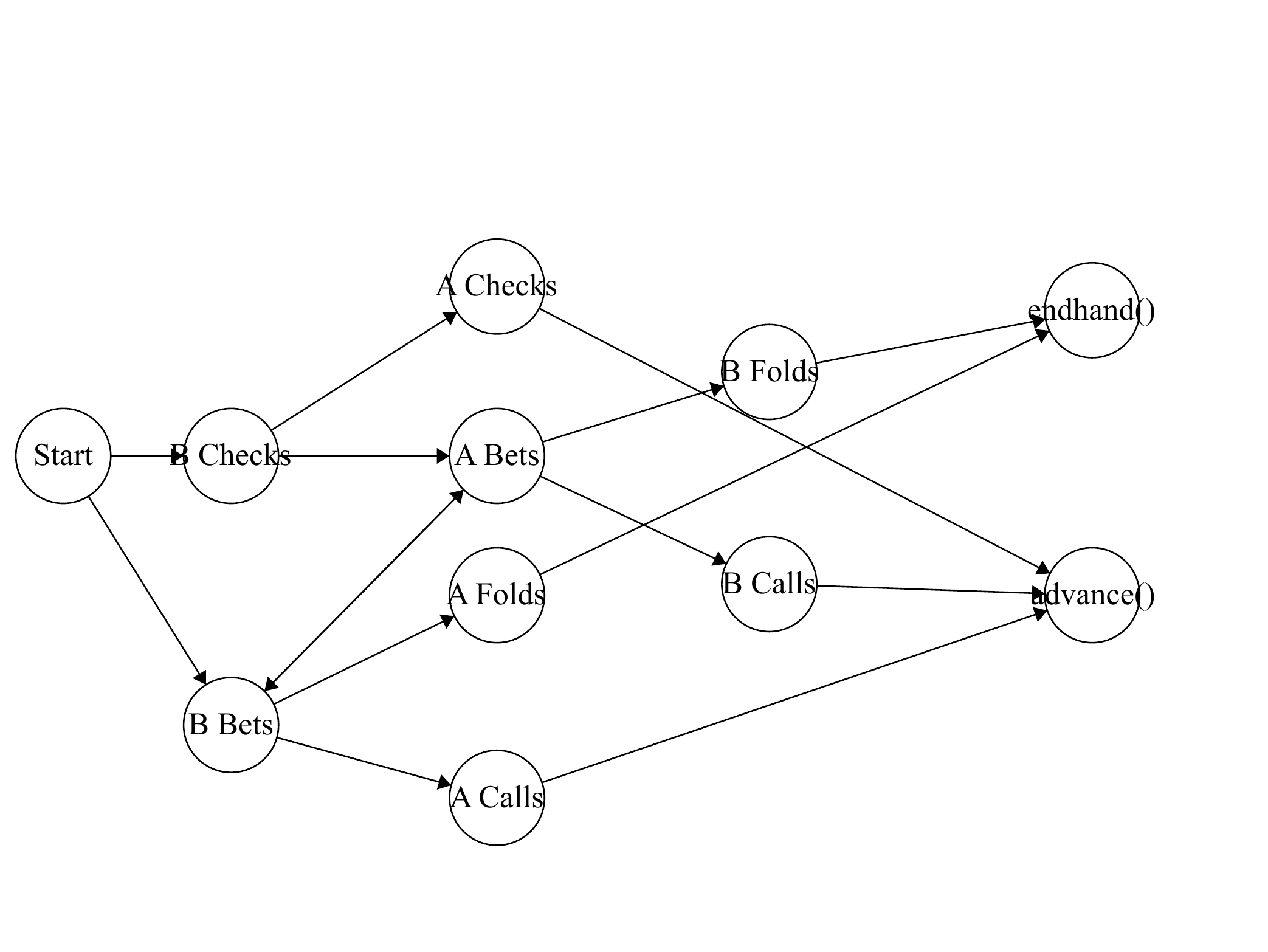}
  \caption{Pre-flop DFA (used before any community cards are dealt).
  \label{fig:dfa1}
}
\end{figure}

\begin{figure}
  \centering
  \includegraphics[width=\linewidth]{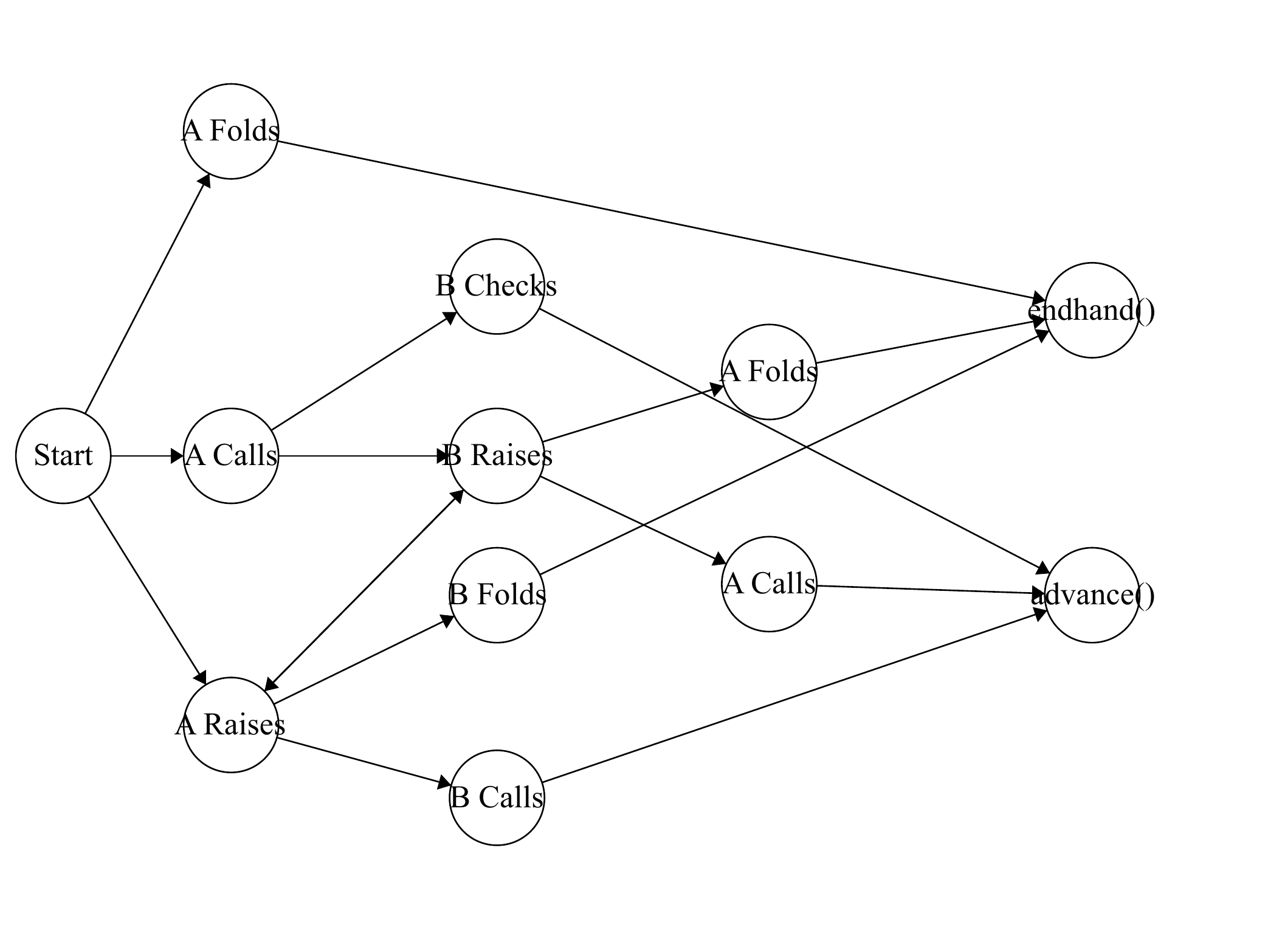}
  \caption{Post-flop DFA (used once community cards have started to been dealt).
  \label{fig:dfa2}
}
\end{figure}

\hspace{\parindent}
The first implementation subproblem to solve was to figure out how to model a game of poker in code. By nature of the rules of poker, the game is deterministic based on previous player actions within a betting round. Thus, the game state can be modeled using a Deterministic Finite Automaton (DFA). The DFA for our application determines the legal actions and/or termination state of the betting round, given previous actions within the betting round. At the end of each betting round, one of two termination states is reached, dictating whether the hand has ended or players will advance to the next betting round. Fig ~\ref{fig:dfa1} and Fig ~\ref{fig:dfa2} show a graphical representation of the DFAs we used in our implementation. In both DFAs, the application begins with the \textit{Start} state and terminates in either the \textit{endHand()} or \textit{advance()} state. The double-headed arrow represents the possible cycle of betting, raising, reraising, etc., until one of the players is eventually all-in. Player 'A' is the one assigned ‘opponent’ at the start of a hand, and player 'B' is the ‘dealer.’


\subsection{Subproblem 2: Rendering 3D Objects Stably}
\hspace{\parindent}
The second implementation subproblem was to figure out how to render 3D objects in the world and effectively track them. Originally the implementation used Snap’s World Tracking \cite{snap_tracking_nodate} functionality, and then its Surface Tracking \cite{snap_tracking_nodate} functionality, with neither proving to be too accurate. Chip stacks are small and users expect them to stay in relatively the same position throughout a game. However, with World Tracking and Surface Tracking, chip stacks would move throughout the room quite a bit if the mobile device’s camera was moved.

Then, we decided to use Marker Tracking \cite{snap_tracking_nodate}, which uses a printed marker pattern to mark a position in the physical world and allow the application to render objects relative to that position. Marker tracking proved to be very accurate and stable for rendering 3D objects in AR. Chip stacks would no longer move throughout the room, as they were grounded in a location in 3D space. Marker Tracking, however, is only a temporary solution while alternative tracking solutions are improved with continued computer vision research.

\subsection{Subproblem 3: Connecting Multiple Players}
\hspace{\parindent}
The third implementation subproblem was to figure out how to connect two players within a single Snapchat Lens to play together and share game data. To solve this problem, we designed an API to connect two different mobile devices and allow them to send messages between each other, including updates on the game state and player actions. This API is built on top of Snap's Connected Lenses feature \cite{snap_connected_nodate}. Thus, the two players will always observe the same data on different devices in real time, unifying their gameplay experience.

\section{Evaluation}
\hspace{\parindent}
We recruited 8 participants who were found through a poker club on campus and recruited by email. We asked participants to respond to a pre-study survey to learn about their individual experience with poker and its rules. This survey can be found in Appendix B. We randomly paired participants into 4 dyads to utilize PokAR to play heads-up poker for 25 minutes. Then, we asked them to respond to a post-study survey about their experience with the application. In this survey, we asked participants about the benefits and detriments of particular PokAR features, the effects of AR on socialization, the effects of AR on game pace, and the overall experience with PokAR. This survey can be found in Appendix C. The study protocol above is described in more detail in Appendix A.


\section{Results}
\hspace{\parindent}
Although participants had varying levels of poker expertise, they all had a self-reported understanding of the rules. Based on the results of the pre-study survey, we grouped the 8 participants into three groups of varying experience levels for analysis: Highly Experienced (play poker multiple times a week, $n=2$), Moderately Experienced (play poker weekly to monthly, $n=3$), and Slightly Experienced (play poker yearly or less, $n=3$).

\subsection{Evaluation of Features}
\hspace{\parindent}
We asked participants to rate each of the five major PokAR features on a scale of 1 (detrimental) to 5 (beneficial) in terms of its effectiveness compared to the corresponding object/action in real-life poker. Each feature earned an average score $> 3$ (leaning beneficial) among all participants. Specifically, "3D AR Chips" scored a 4, "UI Action Buttons" scored a 4.25, "Game Messages" scored a 3.875, "Counting Stacks" scored a 4.375, and "Awarding Pots" scored a 4.25.

Notably, the only features that scored $< 3$ (leaning detrimental) were “3D AR Chips,” “UI Action Buttons,” and “Game Messages” for the Highly Experienced subgroup of participants. This could be explained by the fact that all three of these features are intended to alleviate the requirement of knowing the complex rules of poker. However, in the pre-study survey, all members of this subgroup answered that they play poker quite often and they confidently understand all the rules, so these features likely just got in the way of their gameplay. The features of “Counting Stacks” and “Awarding Pots” were, however, positively received by all three subgroups of participants.

\subsection{Evaluation of Socialization}
\hspace{\parindent}
The average response to the survey question: “How much did AR affect the in-person social aspects of the game of poker?” was a 3.25 on a scale of 1 (negatively) to 5 (positively), meaning that AR neither significantly improved nor impaired the in-person social aspects of poker. This is a beneficial result, as one of PokAR's goals was to supplement the game of poker. We did not implement social-related features intended to improve socialization, but this result supports the claim that the AR features of PokAR did not impair socialization. In other words, players are utilizing PokAR as a tool to enable poker play, which does not get in the way of the traditional social interactions at a poker table.

Additionally, multiple participants noted that AR did not heavily influence socialization. P1 stated that the experience was \textit{“no different, we could still talk and converse,”} and P5 stated that AR \textit{“Didn’t affect socialization because everything was still in person.”}

\subsection{Evaluation of Game Pace}
\hspace{\parindent}
The average response to the survey question: “How did augmented reality affect the game pace of poker?” was a 3.75 on a scale of 1 (slowed the game) to 5 (sped up the game), meaning that AR slightly increased the game pace of poker. In this study, the average game pace was 40.3 hands/hour (67.2 hands/hour for the Highly Experienced subgroup). Comparatively, “A typical live poker game will deal 25-30 per hour,” assuming 9 players \cite{fisk_how_2020}. This section requires additional study, including a control session of each participant group playing traditional poker to compare the game pace with AR poker.

\subsection{Evaluation of Overall Experience}
\hspace{\parindent}
On average, participants rated their overall experience at a 4.5/5, overwhelmingly positive. P2 stated that \textit{“It’s quicker, but annoying to hold the phone up.”} P6 stated that \textit{“It was a different experience which took a little getting used to, but I enjoyed it.”} P4 stated that it \textit{“Felt cool to have the chips tracked for you. Definitely could see myself using it on a camping trip or during traveling.”} P5 stated that \textit{“It was cool, because sometimes I’d like to play poker but sometimes have no chips!”}

\section{Conclusion}
\hspace{\parindent}
Through developing a complete AR application and studying how people utilize it, we have generated three main conclusions.

\subsection{AR Has the Potential to Augment and Simplify Traditional Table Games}
\hspace{\parindent}
After our work throughout this semester, we are confident that AR as a technology can and will be used in the future to augment and simplify traditional table games, like poker. While the technology is currently in a primitive state, it is continually evolving and progressing. Several participants noted that the gameplay of PokAR was clunky since they had to constantly hold their mobile devices up to see the game. However, with improved AR technology, this annoyance will begin to fade away. For example, AR glasses like Spectacles will eliminate the need to hold up a mobile device \cite{snap_spectacles_nodate}.

This study revealed promising results concerning the future potential of AR in games. For instance, the use of AR did not hinder socialization, and participants had a positive overall experience. PokAR features meant to facilitate gameplay were positively received by most players, and options to disable disruptive features would alleviate the rest.

\subsection{AR Should Not Be Used to Replace Traditional Experiences; It Should Be Used to Augment Them}
\hspace{\parindent}
This conclusion stems from the fact that AR technology has innate limitations compared to the physical world. For instance, AR experiences are less tactile than physical world experiences. While software tricks exist to improve the tactility of AR experiences (like hand tracking, which enables object manipulation), it will never feel quite like the physical world. For example, several participants noted that PokAR was missing one important aspect of traditional poker: chip shuffling. Chip shuffling is a common fidgeting technique among poker players in which they use one hand to rearrange a stack of chips. Shuffling is almost unanimous among poker players and is commonly used to pass time and cure boredom during long poker sessions. While AR could simulate chip shuffling, it could never reproduce the experience perfectly.

Early intuition about this conclusion is one reason we decided to utilize physical playing cards in PokAR. If playing cards were virtual, players would be playing an online poker game in which in-person social interactions were minimal. Players wouldn’t even need to be co-located to play PokAR anymore. This would be a case of using AR to replace a traditional game experience. Instead, we decided to use physical cards and virtual chips in PokAR to afford some physical-world tactility to players and streamline some of the more annoying and time-consuming aspects of poker, like counting chips.

As mentioned in the introduction, PokAR is not intended to replace traditional poker, only augment it. This is due to the inherent limitations of AR technology. More broadly, AR should not be used to replace traditional experiences, only augment them. Augmentations should be deliberately planned and carefully implemented to ensure that they do not take over the spirit of the game. Go too far with augmentation, and you approach the virtual reality world and lose out on social interaction. 

\subsection{Future Work}
\hspace{\parindent}
Our work on PokAR has revealed possible directions for future study and enhancements to the application. Firstly, due to time constraints, not all features of poker were able to be added. PokAR is currently limited to just two players. This design choice was made to reduce the project's complexity, but this limit should be increased to the accepted limit of nine players to better simulate traditional poker. Additionally, the option to chop pots (split pots equally between tied players) is currently not implemented. The option to run it multiple times (deal remaining cards multiple times in an all-in situation and award the pot proportionally to winners) is also not yet implemented.

PokAR would benefit from increased tactility, which is why we believe that it is a necessary direction for future work. Increased tactility could come in two forms, with the first being the ability to grab AR chips and manipulate them with your hands. This feature would let players bet more realistically (rather than just clicking a button) or could help simulate chip shuffling (which was mentioned earlier as a lacking aspect). The other way to increase tactility would be to utilize hand gestures, rather than UI buttons, to signal actions. For example, players could tap the table with a fist to signal a ‘check,’ as in traditional poker. These features would further increase the immersion of PokAR.

Finally, AR could be utilized to provide helpful statistical annotations for players. Possible annotations could include the probability of winning or the probability of making a certain hand. To implement this feature, one must first implement a computer vision model to recognize and classify playing cards. This has been done in the past with high accuracy ($> 99$\%) \cite{rohlfing-das_image_2020}. This feature would further reduce the mental load on players and help them play and learn poker more effectively.


\pagebreak

\bibliographystyle{ACM-Reference-Format}
\bibliography{pokar}


\begin{thebibliography}{16}


\ifx \showCODEN    \undefined \def \showCODEN     #1{\unskip}     \fi
\ifx \showDOI      \undefined \def \showDOI       #1{#1}\fi
\ifx \showISBNx    \undefined \def \showISBNx     #1{\unskip}     \fi
\ifx \showISBNxiii \undefined \def \showISBNxiii  #1{\unskip}     \fi
\ifx \showISSN     \undefined \def \showISSN      #1{\unskip}     \fi
\ifx \showLCCN     \undefined \def \showLCCN      #1{\unskip}     \fi
\ifx \shownote     \undefined \def \shownote      #1{#1}          \fi
\ifx \showarticletitle \undefined \def \showarticletitle #1{#1}   \fi
\ifx \showURL      \undefined \def \showURL       {\relax}        \fi
\providecommand\bibfield[2]{#2}
\providecommand\bibinfo[2]{#2}
\providecommand\natexlab[1]{#1}
\providecommand\showeprint[2][]{arXiv:#2}

\bibitem[Dagan et~al\mbox{.}(2022)]%
        {dagan_project_2022}
\bibfield{author}{\bibinfo{person}{Ella Dagan}, \bibinfo{person}{Ana~Cárdenas
  Gasca}, \bibinfo{person}{Ava Robinson}, \bibinfo{person}{Anwar Noriega},
  \bibinfo{person}{Yu~Jiang Tham}, \bibinfo{person}{Rajan Vaish}, {and}
  \bibinfo{person}{Andrés Monroy-Hernández}.}
  \bibinfo{year}{2022}\natexlab{}.
\newblock \showarticletitle{Project {IRL}: {Playful} {Co}-{Located}
  {Interactions} with {Mobile} {Augmented} {Reality}}.
\newblock \bibinfo{journal}{\emph{Proceedings of the ACM on Human-Computer
  Interaction}} \bibinfo{volume}{6}, \bibinfo{number}{CSCW1}
  (\bibinfo{date}{March} \bibinfo{year}{2022}), \bibinfo{pages}{1--27}.
\newblock
\showISSN{2573-0142}
\urldef\tempurl%
\url{https://doi.org/10.1145/3512909}
\showDOI{\tempurl}
\newblock
\shownote{arXiv:2201.02558 [cs]}.


\bibitem[Fisk(2020)]%
        {fisk_how_2020}
\bibfield{author}{\bibinfo{person}{Geoffrey Fisk}.}
  \bibinfo{year}{2020}\natexlab{}.
\newblock \bibinfo{title}{How {Many} {Hands} {Are} {Played} {Per} {Hour} in
  {Live} {Poker} {Games}?}
\newblock
\newblock
\urldef\tempurl%
\url{https://upswingpoker.com/hands-per-hour-live-poker-vs-online/}
\showURL{%
\tempurl}


\bibitem[Herodotou(2010)]%
        {herodotou_social_2010}
\bibfield{author}{\bibinfo{person}{Christothea Herodotou}.}
  \bibinfo{year}{2010}\natexlab{}.
\newblock \showarticletitle{Social {Praxis} {Within} and {Around} {Online}
  {Gaming}: {The} {Case} of {World} of {Warcraft}}. In
  \bibinfo{booktitle}{\emph{2010 {Third} {IEEE} {International} {Conference} on
  {Digital} {Game} and {Intelligent} {Toy} {Enhanced} {Learning}}}.
  \bibinfo{pages}{10--22}.
\newblock


\bibitem[Herodotou et~al\mbox{.}(2015)]%
        {herodotou_iterative_2015}
\bibfield{author}{\bibinfo{person}{Christothea Herodotou},
  \bibinfo{person}{Niall Winters}, {and} \bibinfo{person}{Maria Kambouri}.}
  \bibinfo{year}{2015}\natexlab{}.
\newblock \showarticletitle{An {Iterative}, {Multidisciplinary} {Approach} to
  {Studying} {Digital} {Play} {Motivation}: {The} {Model} of {Game}
  {Motivation}}.
\newblock \bibinfo{journal}{\emph{Games and Culture}} \bibinfo{volume}{10},
  \bibinfo{number}{3} (\bibinfo{date}{May} \bibinfo{year}{2015}),
  \bibinfo{pages}{249--268}.
\newblock
\showISSN{1555-4120, 1555-4139}
\urldef\tempurl%
\url{https://doi.org/10.1177/1555412014557633}
\showDOI{\tempurl}


\bibitem[Lee et~al\mbox{.}(2021)]%
        {lee_social_2021}
\bibfield{author}{\bibinfo{person}{Hyun-Woo Lee}, \bibinfo{person}{Sanghoon
  Kim}, {and} \bibinfo{person}{Jun-Phil Uhm}.} \bibinfo{year}{2021}\natexlab{}.
\newblock \showarticletitle{Social {Virtual} {Reality} ({VR}) {Involvement}
  {Affects} {Depression} {When} {Social} {Connectedness} and {Self}-{Esteem}
  {Are} {Low}: {A} {Moderated} {Mediation} on {Well}-{Being}}.
\newblock \bibinfo{journal}{\emph{Frontiers in Psychology}}
  \bibinfo{volume}{12} (\bibinfo{year}{2021}).
\newblock
\showISSN{1664-1078}
\urldef\tempurl%
\url{https://www.frontiersin.org/articles/10.3389/fpsyg.2021.753019}
\showURL{%
\tempurl}


\bibitem[Meta(2019)]%
        {meta_poker_2019}
\bibfield{author}{\bibinfo{person}{Meta}.} \bibinfo{year}{2019}\natexlab{}.
\newblock \bibinfo{title}{Poker {VR} - {Multi} {Table} {Tournaments} on
  {Oculus} {Quest}}.
\newblock
\newblock
\urldef\tempurl%
\url{https://www.oculus.com/experiences/quest/2257223740990488/}
\showURL{%
\tempurl}


\bibitem[Offshore(2021)]%
        {fast_offshore_online_2021}
\bibfield{author}{\bibinfo{person}{Fast Offshore}.}
  \bibinfo{year}{2021}\natexlab{}.
\newblock \bibinfo{title}{Online poker sector overview for 2021: {Stats}, key
  drivers and more}.
\newblock
\newblock
\urldef\tempurl%
\url{https://fastoffshore.com/2021/10/online-poker-sector-overview-2021/}
\showURL{%
\tempurl}


\bibitem[Rohlfing-Das(2020)]%
        {rohlfing-das_image_2020}
\bibfield{author}{\bibinfo{person}{Arjun Rohlfing-Das}.}
  \bibinfo{year}{2020}\natexlab{}.
\newblock \bibinfo{title}{Image {Classification} for {Playing} {Cards}}.
\newblock
\newblock
\urldef\tempurl%
\url{https://medium.com/swlh/image-classification-for-playing-cards-26d660f3149e}
\showURL{%
\tempurl}


\bibitem[Sakuma et~al\mbox{.}(2012)]%
        {sakuma_enhancing_2012}
\bibfield{author}{\bibinfo{person}{Hiroyuki Sakuma}, \bibinfo{person}{Tetsuo
  Yamabe}, {and} \bibinfo{person}{Tatsuo Nakajima}.}
  \bibinfo{year}{2012}\natexlab{}.
\newblock \showarticletitle{Enhancing {Traditional} {Games} with {Augmented}
  {Reality} {Technologies}}. In \bibinfo{booktitle}{\emph{2012 9th
  {International} {Conference} on {Ubiquitous} {Intelligence} and {Computing}
  and 9th {International} {Conference} on {Autonomic} and {Trusted}
  {Computing}}}. \bibinfo{pages}{822--825}.
\newblock
\urldef\tempurl%
\url{https://doi.org/10.1109/UIC-ATC.2012.95}
\showDOI{\tempurl}


\bibitem[Savela et~al\mbox{.}(2020)]%
        {savela_does_2020}
\bibfield{author}{\bibinfo{person}{Nina Savela}, \bibinfo{person}{Atte
  Oksanen}, \bibinfo{person}{Markus Kaakinen}, \bibinfo{person}{Marius
  Noreikis}, {and} \bibinfo{person}{Yu Xiao}.} \bibinfo{year}{2020}\natexlab{}.
\newblock \showarticletitle{Does {Augmented} {Reality} {Affect} {Sociability},
  {Entertainment}, and {Learning}? {A} {Field} {Experiment}}.
\newblock \bibinfo{journal}{\emph{Applied Sciences}} \bibinfo{volume}{10},
  \bibinfo{number}{4} (\bibinfo{date}{Jan.} \bibinfo{year}{2020}),
  \bibinfo{pages}{1392}.
\newblock
\showISSN{2076-3417}
\urldef\tempurl%
\url{https://doi.org/10.3390/app10041392}
\showDOI{\tempurl}
\newblock
\shownote{Number: 4 Publisher: Multidisciplinary Digital Publishing Institute}.


\bibitem[Snap(2022a)]%
        {snap_connected_nodate}
\bibfield{author}{\bibinfo{person}{Snap}.} \bibinfo{year}{2022}\natexlab{a}.
\newblock \bibinfo{title}{Connected {Lenses} {Overview} {\textbar} {Docs}}.
\newblock
\newblock
\urldef\tempurl%
\url{https://docs.snap.com/lens-studio/references/guides/lens-features/connected-lenses/connected-lenses-overview}
\showURL{%
\tempurl}


\bibitem[Snap(2022b)]%
        {snap_how_nodate}
\bibfield{author}{\bibinfo{person}{Snap}.} \bibinfo{year}{2022}\natexlab{b}.
\newblock \bibinfo{title}{How do {I} use {Lenses} on {Snapchat}?}
\newblock
\newblock
\urldef\tempurl%
\url{https://support.snapchat.com/en-US/a/face-world-lenses}
\showURL{%
\tempurl}


\bibitem[Snap(2022c)]%
        {snap_spectacles_nodate}
\bibfield{author}{\bibinfo{person}{Snap}.} \bibinfo{year}{2022}\natexlab{c}.
\newblock \bibinfo{title}{Spectacles by {Snap} {Inc}. • {The} {Next}
  {Generation} of {Spectacles}}.
\newblock
\newblock
\urldef\tempurl%
\url{https://www.spectacles.com/}
\showURL{%
\tempurl}


\bibitem[Snap(2022d)]%
        {snap_tracking_nodate}
\bibfield{author}{\bibinfo{person}{Snap}.} \bibinfo{year}{2022}\natexlab{d}.
\newblock \bibinfo{title}{Tracking {Modes} {\textbar} {Docs}}.
\newblock
\newblock
\urldef\tempurl%
\url{https://docs.snap.com/lens-studio/references/guides/lens-features/tracking/world/tracking-modes}
\showURL{%
\tempurl}


\bibitem[Thul(2013)]%
        {thul_pokertool_2013}
\bibfield{author}{\bibinfo{person}{Christoph Thul}.}
  \bibinfo{year}{2013}\natexlab{}.
\newblock \showarticletitle{{PokerTool} - {Entwicklung} und {Implementierung}
  einer {AR}-{Android}-{Anwendung} für {Wahrscheinlichkeitsberechnungen} bei
  {Texas} {Holdem} {Poker}}.
\newblock  (\bibinfo{date}{Sept.} \bibinfo{year}{2013}).
\newblock
\urldef\tempurl%
\url{https://kola.opus.hbz-nrw.de/opus45-kola/frontdoor/index/index/docId/769}
\showURL{%
\tempurl}


\bibitem[Weiser(1991)]%
        {weiser_computer_1991}
\bibfield{author}{\bibinfo{person}{Mark Weiser}.}
  \bibinfo{year}{1991}\natexlab{}.
\newblock \showarticletitle{The {Computer} for the 21st {Century}}.
\newblock  (\bibinfo{year}{1991}).
\newblock


\end{thebibliography}

\appendix
\pagebreak

\section{Study Protocol}
Below is the process we asked participants to follow during the study:
\begin{enumerate}
\item Participants were found through a poker club on campus and recruited by email.
\item Participants were asked to respond to the pre-study survey.
\item The participants were randomly paired into dyads for heads-up poker play.
\item During the study:
    \begin{enumerate}
      \item Participants were asked to download Snapchat (if necessary) and scan a code to gain access to the PokAR Snapchat Lens.
      \item Participants were asked to sign consent forms.
      \item Participants were asked to use PokAR to play heads-up poker (without using real-world money) for 25 minutes.
      \item We took notes on comments, reactions, game pace, frustrations, etc. We took photos and videos throughout. We also answered questions about the application when asked, but we avoided guiding the players.
    \end{enumerate}
\item After play, participants were asked to respond to the post-study survey.
\end{enumerate}

\section{Pre-Study Survey}
Below are the questions asked during the pre-study survey.
\begin{itemize}
    \item How well do you know the rules of Heads-Up Texas Hold'em Poker?
    \item How often do you play poker?
\end{itemize}

\section{Post-Study Survey}
Below are the questions asked during the post-study survey.
\begin{itemize}
    \item Please rate each of the following PokAR features in terms of its effectiveness when compared to the corresponding object/action in real-life Texas hold'em poker? 
    \begin{itemize}
        \item 3D AR Chips
        \item UI Action Buttons
        \item Game Messages
        \item Counting Stacks
        \item Awarding Pots
    \end{itemize}
    \item How much did AR affect the in-person social aspects of the game of poker?
    \item How did augmented reality affect the game pace of poker (\# of hands played / unit time)? Ignore the first few hands in which you were learning the application.
    \item Overall, how would you describe your experience with PokAR?
\end{itemize}

\section{Code Repository}
The code for this project can be found at the following GitHub repository: https://github.com/adamgamba/PokAR.

\end{document}